\documentclass[12pt]{article}
\usepackage{graphicx}
\begin{document}

\begin{center}

{\Large Is it possible to construct the Proton Structure Function
by Lorentz-boosting the Static Quark-model Wave Function?} \\[6mm]

Y. S. Kim\footnote{electronic address: yskim@physics.umd.edu}\\
Department of Physics, University of Maryland,\\
College Park, Maryland 20742, U.S.A.\\

\vspace{5mm}

Marilyn E. Noz \footnote{electronic address: noz@nucmed.med.nyu.edu}\\
Department of Radiology, New York University,\\ New York,
New York 10016, U.S.A.

\end{center}

\begin{abstract}
The energy-momentum relations for massive and massless particles
are $E = p^2/2m $ and $E = pc$ respectively.  According to Einstein,
these two different expressions come from the same formula
$E = \sqrt{(cp)^2 + m^2 c^4}$.  Quarks and partons are believed to be
the same particles, but they have quite different properties.  Are
they two different manifestations of the same covariant entity as
in the case of Einstein's  energy-momentum relation?  The answer to
this question is YES.  It is possible to construct harmonic oscillator
wave functions which can be Lorentz-boosted.  They describe quarks
bound together inside hadrons.  When they are boosted to an
infinite-momentum frame, these wave functions exhibit all the peculiar
properties of Feynman's parton picture.  This formalism leads to a
parton distribution corresponding to the valence quarks, with a good
agreement with the experimentally observed distribution.
\end{abstract}

\newpage

\section{Introduction}\label{intro}
Since Einstein's formulation of special relativity in 1905, the most
significant addition to physics was quantum mechanics based on
wave-particle duality.  The question then is whether relativity is
consistent with quantum mechanics.  The answer to this question is
very simple for plane waves, because they can can be written in a
Lorentz-invariant form.  Because of this mathematical simplicity,
it is possible to construct quantum field theory with Feynman diagrams
as a computational tool.

How about standing waves?  Are they consistent with Einstein's
relativity?  This question has not been properly addressed.
The question is how standing waves in one Lorentz frame would look
in a different Lorentz frame.  While we are not able to
address this problem in general terms, we would like to point out
that there is at least one set of wave functions which can be
Lorentz-boosted.  It is the set of solutions of the Lorentz-invariant
harmonic oscillator equation proposed by Feynman {\it et al.} in
1971~\cite{fkr71}.  The solutions given in their papers are
not normalizable in time-separation variable and do not correspond
to physics.

However, there are more than two hundred solutions satisfying different
boundary conditions.  We choose the solutions which are localized and
normalizable in both space and time coordinates. We shall call this set
of solutions covariant harmonic oscillators.  These solutions have the
following common properties:

\begin{itemize}

\item[1).]  The formalism is consistent with established physical
principles including the uncertainty principle in quantum mechanics
and the transformation laws of special relativity~\cite{knp86}.

\item[2).] The formalism is consistent with the basic hadronic
features observed in high-energy laboratories, including hadronic mass
spectra, the proton form factor, and the parton phenomena~\cite{knp86}.

\item[3).] The formalism constitutes a representation of Wigner's little
of the Poincar\'e group which dictates internal space-time symmetries of
relativistic particles~\cite{knp86,wig39}.  The little group is the
maximal subgroup of the Lorentz group which leaves the four-momentum
of a given particle invariant.

\end{itemize}

In this paper, we use this covariant oscillator formalism to see that
the quark model and the parton models are two different manifestations
of one covariant formalism.  We shall see how the parton picture
emerges from the Lorentz-boosted hadronic wave function. In
Sec.~\ref{covham}, we introduce the basic ingredients of the covariant
harmonic oscillator formalism.  In Sec.~\ref{parton}, we use this
formalism to show that the valance quark distribution in the proton
structure function can be derived from the Lorentz-boosted quark-model
wave function.

\section{Covariant Harmonic Oscillators}\label{covham}
The covariant harmonic oscillator formalism has been discussed
exhaustively in the literature, and it is not necessary to give another
full-fledged treatment in the present paper.  We shall discuss here
only its features needed for explaining the peculiarities of Feynman's
parton picture.

\begin{figure}[thb]
\centerline{\includegraphics[scale=0.6]{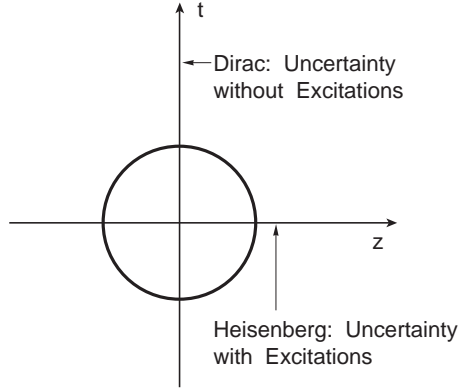}}
\caption{Quantum mechanics with the c-number time-energy uncertainty
relation.  The present form of quantum mechanics allows quantum
excitations along the space-like directions, but does not allow
excitations along the time-like direction even though there is an
uncertainty relation between the time and energy variables.}\label{quantum}

\end{figure}

Let us consider a bound state of two particles.  Then there is a
Bohr-like radius measuring the space-like separation
between the quarks.  There is also a time-like separation between the
quarks, and this variable becomes mixed with the longitudinal spatial
separation as the hadron moves with a relativistic speed.  There are
no quantum excitations along the time-like direction.  On the other
hand, there is the time-energy uncertainty relation which allows quantum
transitions.  The covariant oscillator formalism can accommodate
these aspects within the framework of the present form of quantum mechanics.
The uncertainty
relation between the time and energy variables is the c-number
relation~\cite{dir27}, which does not allow excitations along the
time-like coordinate.  This aspect is illustrated in Fig.~\ref{quantum}.

Let us consider now a hadron consisting of two quarks.  If the space-time
position of two quarks are specified by $x_{a}$ and $x_{b}$ respectively,
the system can be described by the variables
\begin{equation}
X = (x_{a} + x_{b})/2 , \qquad x = (x_{a} - x_{b})/2\sqrt{2} .
\end{equation}
The four-vector $X$ specifies where the hadron is located in space and
time, while the variable $x$ measures the space-time separation between
the quarks.  In the convention of Feynman {\it et al.}~\cite{fkr71}, the
internal motion of the quarks bound by a harmonic oscillator potential
of unit strength can be described by the Lorentz-invariant equation
\begin{equation}\label{feyn71}
\frac{1}{2}\left\{x^{2}_{\mu } -
\frac{\partial ^{2}}{\partial x_{\mu }^{2}} \right\} \psi (x)
= \lambda \psi (x) .
\end{equation}
We use here the metric: $x^{\mu } = (x, y, z, t)$.

If the hadron is at rest, we can consider a solution of the form
\begin{equation}
\psi (x,y,z,t) = \psi (x,y,z) \left(\frac{1}{\pi}\right)^{1/4}
\exp{\left(-t^{2}/2 \right)} ,
\end{equation}
where $\psi (x,y,z)$ is the wave function for the three-dimensional
oscillator with appropriate angular momentum quantum numbers.  Indeed,
the above wave function constitutes a representation of Wigner's
$O(3)$-like little group for a massive particle~\cite{knp86}.  In the
above expression, there are no time-like excitations, and this is
consistent with what we see in the real world.

\begin{figure}[thb]
\centerline{\includegraphics[scale=0.6]{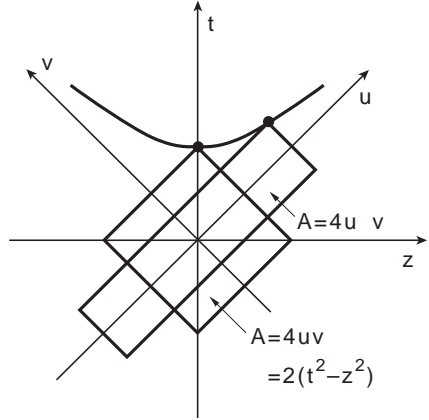}}
\caption{Space-time picture of the Lorentz boost.  The invariant
quantity $(z^{2} - t^{2})$ can be written as $(z + t)(z - t)$.  This
is proportional to the product of the light-cone variables
$u = (z + t)/\sqrt{2}$ and $(z - t)/\sqrt{2}$.}\label{licone}
\end{figure}

Since the three-dimensional oscillator differential equation is
separable in both spherical and Cartesian coordinate systems,
$\psi (x,y,z)$ consists of Hermite polynomials of $x, y$, and $z$.  If
the Lorentz boost is made along the $z$ direction, the $x$ and $y$
coordinates are not affected, and can be dropped from the wave function.
The wave function of interest can be written as
\begin{equation}
\psi^{n}(z,t) = \left(\frac{1}{\pi}\right)^{1/4}
\exp{\left(-t^{2}/2\right)}\psi_{n}(z) ,
\end{equation}
with
\begin{equation}
\psi ^{n}(z) = \left(\frac{1}{\pi n!2^{n}} \right)^{1/2} H_{n}(z)
\exp (-z^{2}/2) ,
\end{equation}
where $\psi ^{n}(z)$ is for the $n$-th excited oscillator state.
The full wave function $\psi ^{n}(z,t)$ is
\begin{equation}\label{wf1}
\psi ^{n}_{0}(z,t) = \left(\frac{1}{\pi n! 2^{n}}\right)^{1/2} H_{n}(z)
\exp \left\{-\frac{1}{2}\left(z^{2} + t^{2} \right) \right\} .
\end{equation}
The subscript $0$ means that the wave function is for the hadron at rest.
The above expression is not Lorentz-invariant, and its localization
a deformation as the hadron moves along the $z$ direction~\cite{knp86}.
This is still a Lorentz-covariant expression, and this form satisfies the
Lorentz-invariant differential equation of Eq.(\ref{feyn71}) even if
the $z$ and $t$ variables are given by
\begin{equation}\label{boost}
(\cosh\eta)z - (\sinh\eta)t,  \qquad
(\cosh\eta)t - (\sinh\eta)z.  \qquad
\end{equation}
This corresponds to a Lorentz-boosting of the system along the $z$
direction with the boost parameter $\eta.$   This becomes
more transparent if we use the light-cone if we use the light-cone
coordinate system where
\begin{equation}
u = (z + t)/\sqrt{2}, \qquad  v = (z - t)/\sqrt{2},
\end{equation}
as is illustrated in Fig.~\ref{licone}.  Here one coordinate is becoming
expanded while the other become contracted.  This type of deformation is
called ``squeeze'' these days~\cite{knp91},

The wave function becomes
\begin{equation}\label{wf2}
\psi_{0}(z,t) = \left(\frac{1}{\pi} \right)^{1/2}
\exp \left\{-\frac{1}{2} (u^{2} + v^{2}) \right\} ,
\end{equation}
where we have left out the Hermite polynomials for simplicity, because
the essential properties of the oscillator wave functions are dominated
by their Gaussian factor.

\begin{figure}[thb]
\centerline{\includegraphics[scale=0.4]{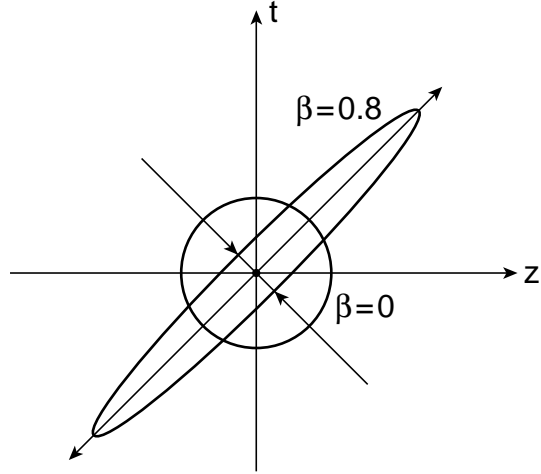}}
\caption{Effect of the Lorentz boost on the space-time wave function.
The circular space-time distribution at the rest frame becomes
Lorentz-squeezed to become an elliptic distribution.}\label{ellipse}
\end{figure}

If the system is boosted variables $u$ and $v$ are replaced by
$e^{-\eta}u$ and $e^{eta} v$ respectively, as is illustrated
in Fig.~\ref{licone}.
and the Lorentz-squeezed wave function becomes
\begin{equation}\label{wf3}
\psi _{\eta }(z,t) = \left(\frac{1}{\pi}\right)^{1/2}
\exp \left\{-\frac{1}{2}\left(e^{-2\eta }u^{2} +
e^{2\eta }v^{2}\right)\right\} .
\end{equation}
The transition from Eq.(\ref{wf2}) to Eq.(\ref{wf3}) is illustrated
in Fig.~\ref{ellipse}.  We can produce this figure by combining
quantum mechanics of Fig.~\ref{quantum} and special relativity
of Fig.~\ref{licone}.

\section{Feynman's Parton Picture}\label{partonp}
In 1969~\cite{fey69}, Feynman observed the following peculiarities in
his parton picture of hadrons.

\begin{itemize}
\item[1).] The picture is valid only for hadrons moving with velocity close
       to that of light.

\item[2).] The interaction time between the quarks becomes dilated, and
        partons behave as free independent particles.

\item[3).] The momentum distribution of partons becomes widespread as the
       hadron moves fast.

\item[4).] The number of partons seems to be infinite or much larger than
       that of quarks.

\end{itemize}

\noindent Because the hadron is believed to be a bound state of two or three
quarks, each of the above phenomena appears as a paradox, particularly 2) and
3) together.  We would like to resolve this paradox using the covariant
harmonic oscillator formalism.

For this purpose, we need a momentum-energy wave function.  If the quarks
have the four-momenta $p_{a}$ and $p_{b}$, we can construct two independent
four-momentum variables~\cite{fkr71}:
\begin{equation}
P = p_{a} + p_{b} , \qquad q = \sqrt{2}(p_{a} - p_{b}) .
\end{equation}
The four-momentum $P$ is the total four-momentum and is thus the hadronic
four-momentum.  $q$ measures the four-momentum separation between the quarks.

\begin{figure}
\centerline{\includegraphics[scale=0.5]{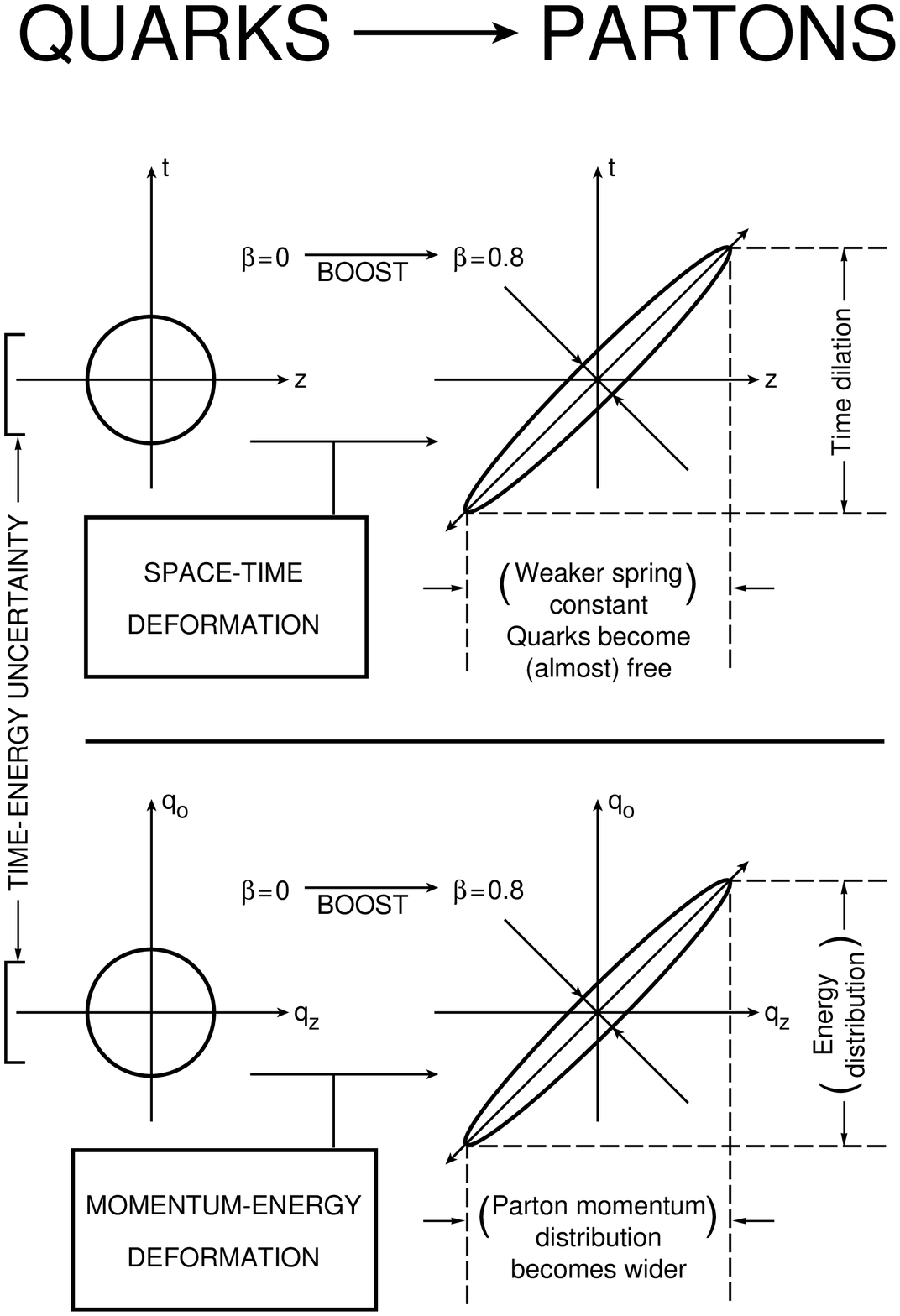}}
\caption{Lorentz-squeezed space-time and momentum-energy wave
functions.  As the hadron's speed approaches that of light, both wave
function become concentrated along their respective positive
light-cone axes.  These light-cone concentrations lead to Feynman's
parton picture.}\label{parton}

\end{figure}

We expect to get the momentum-energy wave function by taking the Fourier
transformation of Eq.(\ref{wf3}):
\begin{equation}\label{fourier}
\phi_{\eta }(q_{z}, q_{0}) = \left(\frac{1}{2\pi}\right)
\int \psi_{\eta}(z, t) \exp{\left\{-i(q_{z}z - q_{0}t)\right\}} dx dt .
\end{equation}
Let us now define the momentum-energy variables in the light-cone coordinate
system as
\begin{equation}\label{conju}
q_{u} = (q_{0} - q_{z})/\sqrt{2} ,  \qquad
q_{v} = (q_{0} + q_{z})/\sqrt{2} .
\end{equation}
In terms of these variables, the Fourier transformation of
Eq.(\ref{fourier}) can be written as
\begin{equation}\label{fourier2}
\phi_{\eta }(q_{z},q_{0}) = \left(\frac{1}{2\pi}\right)
\int \psi_{\eta}(z, t) \exp{\left\{-i(q_{u} u + q_{v} v)\right\}} du dv .
\end{equation}
The resulting momentum-energy wave function is
\begin{equation}\label{phi}
\phi_{\eta }(q_{z},q_{0}) = \left(\frac{1}{\pi}\right)^{1/2}
\exp\left\{-\frac{1}{2}\left(e^{-2\eta}q_{u}^{2} +
e^{2\eta}q_{v}^{2}\right)\right\} .
\end{equation}
Because we are using here the harmonic oscillator, the mathematical form
of the above momentum-energy wave function is identical to that of the
space-time wave function.  The Lorentz squeeze properties of these wave
functions are also the same, as is indicated in Fig.~\ref{parton}.

When the hadron is at rest with $\eta = 0$, both wave functions behave
like those for the static bound state of quarks.  As $\eta$ increases,
the wave functions become continuously squeezed until they become
concentrated along their respective positive light-cone axes.  Let us
look at the z-axis projection of the space-time wave function.  Indeed,
the width of the quark distribution increases as the hadronic speed
approaches that of the speed of light.  The position of each quark
appears widespread to the observer in the laboratory frame, and the
quarks appear like free particles.

Furthermore, interaction time of the quarks among themselves become
dilated.  Because the wave function becomes wide-spread, the distance
between one end of the harmonic oscillator well and the other end
increases as is indicated in Fig.~\ref{ellipse}.  This effect, first
noted by Feynman~\cite{fey69}, is universally observed in high-energy
hadronic experiments.  Let us look at the time ratio more carefully.
The period of oscillation increases like $e^{\eta}$ as was predicted by
Feynman~\cite{fey69}.

In the picture of the Lorentz
squeezed hadron given in Fig.~\ref{ellipse}, the hadron moves along the $u$
(positive light-cone) axis, while the external signal moves in the
direction opposite to the hadronic momentum, which corresponds to
the $v$ (negative light-cone) axis.  This time interval is proportional
to the minor axis of the ellipse given in Fig.~\ref{ellipse}.

If we use $T_{ext}$ and $T_{osc}$ for the quark's interaction time with
the external signal and the interaction time among the quarks, their ratio
becomes
\begin{equation}\label{ratio}
\frac{T_{ext}}{T_{osc}} = \frac{\exp(-\eta)}{\exp(\eta)} = \exp(-2\eta) .
\end{equation}
The ratio of the interaction time to the oscillator period becomes
$e^{-2\eta}$.  The energy of each proton coming out of the Fermilab
accelerator is $900 GeV$.  This leads to the ratio to $10^{-6}$.  This
is indeed a small number.  The external signal is not able to sense the
interaction of the quarks among themselves inside the hadron.  Thus, the
quarks appear to be free particles to the external signal.
This is the cause of incoherence in the parton interaction amplitudes.
The momentum-energy wave function is just like the space-time wave
function in the oscillator formalism.  The longitudinal momentum
distribution becomes wide-spread as the hadronic speed approaches the
velocity of light.  This is in contradiction with our expectation from
nonrelativistic quantum mechanics that the width of the momentum
distribution is inversely proportional to that of the position wave
function.  Our expectation is that if the quarks are free, they must
have their sharply defined momenta, not a wide-spread distribution.
This apparent contradiction presents to us the following two fundamental
questions:

\begin{itemize}

\item[1).]  If both the spatial and momentum distributions become
      widespread as the hadron moves, and if we insist on Heisenberg's
      uncertainty relation, is Planck's constant dependent on the
      hadronic velocity?

\item[2).]  Is this apparent contradiction related to another apparent
      contradiction that the number of partons is infinite while there
      are only two or three quarks inside the hadron?
\end{itemize}

The answer to the first question is ``No'', and that for the second
question is ``Yes''.  Let us answer the first question which is related
to the Lorentz invariance of Planck's constant.  If we take the product
of the width of the longitudinal momentum distribution and that of the
spatial distribution, we end up with the relation
\begin{equation}
<z^{2}><q_{z}^{2}> = (1/4)[\cosh(2\eta)]^{2}  .
\end{equation}
The right-hand side increases as the velocity parameter increases.  This
could lead us to an erroneous conclusion that Planck's constant becomes
dependent on velocity.  This is not correct, because the longitudinal
momentum variable $q_{z}$ is no longer conjugate to the longitudinal
position variable when the hadron moves.
\begin{figure}[thb]
\centerline{\includegraphics[scale=0.4]{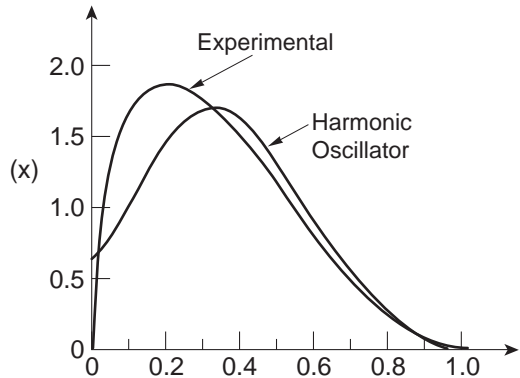}}
\caption{Calculation of the parton distribution from a hadronic wave
function in the static quark model.  The covariant harmonic oscillator
formalism produces the quark-model wave function for hadrons at rest.
The same formalism produces the parton distribution function in the
infinite-momentum frame.  The valence parton distribution calculated
in this way is compared with the distribution observed in high-energy
laboratories. This graph shows a good agreement between the
oscillator-based theory and the observed experimental data.}\label{hussar}

\end{figure}

In order to maintain the Lorentz-invariance of the uncertainty product,
we have to work with a conjugate pair of variables whose product does
not depend on the velocity parameter.  Let us go back to Eq.(\ref{conju})
and Eq.(\ref{fourier2}).  It is quite clear that the light-cone variable
$u$ and $v$ are conjugate to $q_{u}$ and $q_{v}$ respectively.  It is
also clear that the distribution along the $q_{u}$ axis shrinks as the
$u$-axis distribution expands.  The exact calculation leads to
\begin{equation}
<u^{2}><q_{u}^{2}> = 1/4 , \qquad  <v^{2}><q_{v}^{2}> = 1/4  .
\end{equation}
Planck's constant is indeed Lorentz-invariant.

Let us next resolve the puzzle of why the number of partons appears to
be infinite while there are only a finite number of quarks inside the
hadron.  As the hadronic speed approaches the speed of light, both the
$x$ and $q$ distributions become concentrated along the positive
light-cone axis.  This means that the quarks also move with velocity
very close to that of light. Quarks in this case behave like massless
particles.

We then know from statistical mechanics that the number of massless
particles  is not a conserved quantity.  For instance, in black-body
radiation, free light-like particles have a widespread momentum
distribution.  However, this does not contradict the known principles
of quantum mechanics, because the massless photons can be divided into
infinitely many massless particles with a continuous momentum
distribution.

Likewise, in the parton picture, massless free quarks have a wide-spread
momentum distribution.  They can appear as a distribution of an
infinite number of free particles.  These free massless particles are
the partons.  It is possible to measure this distribution in high-energy
laboratories, and it is also possible to calculate it using the covariant
harmonic oscillator formalism.  We are thus forced to compare these two
results~\cite{hussar81}.  Figure~\ref{hussar} shows the result.

\section*{Concluding Remarks}
In this report, we introduced first the covariant harmonic oscillator
formalism which is consistent with all physical laws of quantum mechanics
and special relativity.  We then used this this formalism to show that
the quark model and the parton model are two limiting case of one
covariant picture of quantum bound states.

\end{document}